# Medium Temperature Phase Change Materials Thermal Characterization by the T-History Method and Differential Scanning Calorimetry

D. Gaona[1*], E. Urresta[1], J. Martínez[1,2], G. Guerrón[1].
[1] National Institute of Energy Efficiency and Renewable Energies (INER), Iñaquito N35-37 and Juan Pablo Sanz, Quito, Ecuador, E-mail address: david.gaona@iner.gob.ec,
[2] Facultad de Arquitectura en Ingeniería, Universidad Internacional SEK, 170302 Quito, Ecuador. E-mail address: javier.martinez@uisek.edu.ec

## ABSTRACT

This paper presents a thermal characterization of salt mixtures applying the T-History Method and the Differential Scanning Calorimetry (DSC) techniques. By using water as a standard substance, the original T-History method was developed to analyze materials with melting points under 100°C. This is the first research that proposes to replace water by glycerin to characterize medium melting temperature PCMs (140-220°C). Moreover, the DSC technique was used to validate and compare the results obtained with the T-history method for each mixture. For instance, the system compound of 40% $KNO_3$ – 60% $NaNO_3$ was studied, and the enthalpy of fusion determined by THM and DSC differs by 2.1% between each method. The results given by the two methods for all mixtures showed that both techniques are complementary and present satisfactory agreement with the specialized literature.

**Keywords:** Phase change material, Energy storage, Thermal characterization, T-History method, Differential Scanning Calorimetry.

**Nomenclature**

*Symbol and units*

| | | | |
|---|---|---|---|
| $T_o$ | Initial temperature of PCM, °C | $T_a$ | Ambient temperature, °C |
| $m_p$ | Mass of PCM sample, kg | $\rho_s$ | Density of solid PCM, kg/m$^3$ |
| $Cp_l$ | Liquid heat capacity, J/(kg·K) | $Cp_r$ | Heat capacity of reference substance, J/(kg·K) |
| $Cp_t$ | Heat capacity of sample tube, J/(kg·K) | $r_i$ | Inner radius of sample tube, m |
| $Cp_s$ | Solid heat capacity, J/(kg·K) | $m_r$ | Mass of reference substance, kg |
| $T_m$ | Fusion temperature, °C | $T_f$ | Offset temperature of cooling process, °C |
| $H_m$ | Enthalpy of fusion, J/kg | $T_{m2}$ | Offset temperature of melting process, °C |
| $m_t$ | Mass of sample tube, kg | $r_o$ | Outer radius of sample tube, m |
| $k_s$ | Solid Thermal conductivity, W/(m·K) | $t_s$ | Solidification time of the PCM sample, s |
| $h$ | Natural convection heat transfer coefficient, W/(m$^2$·K) | $Ste_s$ | Stefan Number of solid PCM, $Cp_s (T_m-T_w)/H_m$ |
| $A_c$ | Outer area of sample tube, m$^2$ | $\alpha_s$ | Thermal diffusivity of solid PCM, $k_s/(\rho_s\, Cp_s)$ |
| $T_{m1}$ | Onset temperature of melting process, °C | $T_w$ | Water temperature, °C |

# 1. INTRODUCTION

Increase in energy consumption has been one of the main characteristics throughout the last years. Greenhouse gas emissions have grown significantly due to the big breach between energy consumption and supply. In addition, the continuing rise in energy prices for energy carriers has required researchers to develop new systems that allow using energy efficiently in the daily human activities. In this way, thermal energy storage (TES) plays a significant role in the efficient use of energy field, and a suitable option of this is the capture of the remaining heat in power generation processes. Heat capture and storage are important in the recovery of waste heat from industrial processes, but its efficiency primarily depends on the type of the system that it would be used, and the characteristics of the system. One major characteristic of a storage system is the volumetric heat capacity of the storing substance. It is desirable to employ the minimum volume possible in order to retain the highest amount of thermal energy. Therefore, a proper system should give the largest thermal inertia as possible to the system [1]. The heat storage system and the amount of stored energy depend on the properties of the substances to retain thermal energy either as sensible heat, latent heat or a combination of both cases.

Latent heat implies larger amounts of energy than sensible heat; hence, using energy storage systems based on phase change have resulted attractive to industry. Materials which are used under this principle are denominated phase change materials (PCM). Storage systems utilizing PCMs can be reduced in volume in comparison with sensible heat systems. Nevertheless, heat-transfer design and media selection present larger difficulties [2]. Melting point, density, heat capacity, and enthalpy of fusion are the most important parameters for an energy storage system based on PCMs[3]. Determining these properties is critical for the process of selecting the most suitable materials which can be used as energy storage media. Generally, characterization of materials uses thermal analytic methods to study these properties, for example, the differential scanning calorimetry (DSC) which is a calorimetric method, and the T-history method (THM) [4] that is based on the record of temperature variation as a function of time in a material as it cools down from its melting point to room temperature.

Calorimetry is a thermal analysis technique substantiated on measuring the heat flow generated due to either a chemical reaction or a phase transition process of an element. In both processes, heat can be released or absorbed causing a notable temperature change in the substance. Consequently, calorimetric methods measure the existing heat flow difference between two samples that are maintained at same temperature during a phase transition [5] . In the DSC technique, the temperature of both substances sample and reference are increased at a constant heat flow rate to keep them equal. As DSC operates at a constant pressure, the generated heat flow is analogous to the enthalpy change expressed in mcal $s^{-1}$.

On the other hand, the THM is a powerful tool in the thermal analysis field especially for PCMs because it offers trustworthy results about the way PCMs will behave under a real application [6]. The principal characteristic of this method is it uses considerable quantities of sample to be analyzed and water as reference material, for it eases the macroscopic analysis of material's properties with a melting point under 100°C to be carried out. For PCMs with melting points higher than 100°C, it is necessary to replace water by any other substance with well-known physical and thermal properties.

Comparisons between both methods have been carried out to determine the thermophysical properties of phase change materials with lower melting points, as it can be seen in [7]. According to it, PCMs were characterized through these techniques to establish their availability of being used in photovoltaic applications. The current document presents an analysis of the thermal properties of eutectic nitrate mixtures of sodium nitrate, potassium nitrate and sodium nitrite. These mixes were analyzed by applying the differential scanning calorimetry (DSC) and the T-history method (THM), which was modified from the original proposal to make it valid for medium temperature PCMs with melting points up 100°C

# 2. METHODOLOGY

PCMs were studied by measuring (i) specific heat of solid and liquid phase ($Cp_s$ and $Cp_l$), (ii) melting temperature ($T_m$), and (iii) phase change enthalpy ($H_m$), being $H_m$ and $T_m$ the most valuable properties. These materials were characterized using the T-History Method and Differential Scanning Calorimetry (DSC) to determine their thermophysical properties and compare them with available specialized literature, as it can be seen in table 2. It is important to mention that the literatures listed in Table 2 determine the values of Tm, Hm, $Cp_s$, and $Cp_l$ by using high precision DSC equipment. For instance, [9] establishes that it has been used a DSC with an argon atmosphere to obtain Hm, and Cp was determined by using sapphire. Additionally, [10] remarks it has been used a high sensitivity DSC with aluminum crucibles to analyze PCMs and determine Hm and Cp.

Four systems were selected for evaluation, comprising salts mixtures of sodium nitrate ($NaNO_3$), potassium nitrate ($KNO_3$), and sodium nitrite ($NaNO_2$) in different compositions to formulate binary and ternary mixtures, which are detailed in table 1. These materials were selected because the resulting melting points of the formed eutectic systems are ideal for medium temperature range applications. For DSC testing, a small size sample in the range of 3-5 mg was used while in the THM, it was used a large sample size in the range of 50-100 g of material. The purpose of using large sample sizes in THM technique is to reach a thermodynamic equilibrium and obtain results that can be closer to the real bulk properties. The studied mixtures were prepared according to the composition showed in Table1 and dried for three hours in a stove at 70°C. Next, they were placed in a silica gel drier in order to eliminate residual water that can disturb the results. Samples for DSC analysis were performed under complete temperature cycling experiments. First, mixtures were heated to a 10 K/min temperature rate in DSC equipment until reaching their melting point to obtain homogeneous solutions; subsequently they were cooled at room temperature (20-25°C) to let them solidify as eutectic solids inside the crucibles. Following, a second heating cycle from room temperature to melting point was performed to analyze all its properties. In this way, it can be ensuring that samples are homogeneous in composition at solid state, which is extremely important for obtaining accurate results due to the small amount of sample this method uses.

### 2.1. T-History method

Characterization by THM was performed through heating PCMs samples contained in glass tubes until overpassing their melting point; simultaneously, a tube of glass containing glycerin as reference substance was heated to a similar temperature without reaching its boiling point. Glycerin was used instead of water as the original method proposes because its well-known properties allow the analysis of high melting temperature materials. Both tubes were heated and stabilized at 240°C , for later being cooled down at ambient temperature ($T_a$). Additionally, variation temperature data of PCMs samples and glycerin were recorded and plotted as a function of time (T vs t) [20] for mathematic analysis. The selected PCMs for this study present non-sub cooling THM curves, for the selection of mathematical equations used in this analysis takes into account this particularity. The THM is valid when it satisfies the condition of Biot number $(h \cdot r_o/2k) \leq 0.1$, so it can be ensure there is a uniform temperature distribution all around the sample [21]. Melted samples were placed in 17 cm long glass tubes with 1 cm internal diameter and 0.8mm wall thickness that allows the condition of Biot number to be satisfied. By fulfilling this condition, it can be established that heat transfer occurs only in one dimension along the length of the tube, for lumped capacitance method can be applied [7]. Consequently, it is settled by energy balance that:

$$(m_t Cp_t + m_p C_{pl})(T_o - T_{m1}) = hA_c A_1 \quad (1)$$

$$m_p H_m = hA_c A_2 - (m_t C_{pt})(T_{m1} - T_{m2}) \quad (2)$$

where $A_1 = \int_0^{t_1}(T - T_a)\,dt$, and $A_2 = \int_{t_1}^{t_2}(T - T_a)dt$.. The solid state zone is defined by the equation:

$$(m_t C_{pt} + m_p C_{ps})(T_{m2} - T_f) = hA_c A_3 \quad (3)$$

where $A_3 = \int_{t_2}^{t_3}(T - T_a)dt$.. Considering a Biot number < 0.1, it can be stated that:

$$(m_t Cp_t + m_r C_{pr})(T_o - T_{m1}) = hA_c A'_1 \quad (4)$$

$$(m_tCp_t + m_rC_{pr})(T_{m2} - T_f) = hA_cA'_2 \quad (5)$$

where $A'_1=\int_0^{t'_1}(T - T_a)dt$ and $A_2=\int_{t'_2}^{t'_3}(T - T_a)dt$. From equation (1) to (5), it can be deduced:

$$Cp_s = \frac{m_rCp_r + m_tCp_t}{m_p} \cdot \frac{A_3}{A'_2} - \frac{m_t}{m_p} Cp_t \quad (6)$$

$$Cp_l = \frac{m_rCp_r + m_tCp_t}{m_p} \cdot \frac{A_1}{A'_1} - \frac{m_t}{m_p} Cp_t \quad (7)$$

$$H_m = \frac{m_rCp_r + m_tCp_t}{m_p} \cdot \frac{A_2}{A'_1}(T_0 - T_{m1}) - \frac{mCp(T_{m1} - T_{m2})}{m_p} \quad (8)$$

Knowing the temperature $Tm_2$ becomes a difficult task because the transition toward solid state is a progressive and slow process, and there is not remarkable evidence of a radical temperature change. This means that there is not a peak curve with negative slope that shows a solidifying process. Thereby, it was calculated the inflection point from the THM curve to determine the limit temperature ($Tm_2$) between the phase change process and the solid state. The inflection point corresponds to the minimum value of the first derivative of the curve[22, 23]. As mentioned before, glycerin was used as reference substance (boiling point 290°C) because the analyzed mixtures own high melting points; besides, their thermo physical properties are amply recognized in the specialized literature [24], [25].
Due to glass tubes are slim, which satisfies the condition $D_o < \delta_T$ [26], $\bar{h}$ coefficient can be calculated using the following equation:

$$Nu_L = \frac{4}{3}\left[\frac{7Ra_LPr}{5(20 + 21Pr)}\right]^{0.25} + \frac{4(272 + 315Pr)}{35(64 + 63Pr)}\frac{L}{D_o} \quad (9)$$

## 2.2. Differential Scanning Calorimetry (DSC)

A Metler Toledo high pressure DSC was used, and the samples were placed into 40 uL aluminum crucibles hermetically sealed. DSC was calibrated using Zinc (Zn) and Indium (In) as reference substances, which properties such as melting temperature and fusion enthalpy are widely known in the thermal analysis field:

- Indium: $T_m$= 156.65°C, $H_m$= 28.44 J/g
- Zinc: $T_m$= 419.49°C, $H_m$=107.03 J/g

PCMs samples' weight was kept between 5-7 mg as the technique recommends, and it was established a heating rate of 5-10 K/min. Mixtures were heated in the DSC at 10 K/min rate to achieve total melting of the material to be analyzed, so a homogenous solution was obtained. Afterwards, mixtures in liquid state were cooled down to room temperature (25°C) to ensure good thermal contact between the samples and crucibles, and in this way solid mixtures have uniform composition. Once the samples were melted and cooled to solidify them, the thermal analysis was carried out to determine their properties at a heating rate of 10 k/min. By doing this, errors related to composition can be minimized, and the reliability of results can be settled [27]. Moreover, tests were carried out under an argon atmosphere to avoid chemical reactions at high temperatures because of oxygen presence. Finally, the mixtures were heated to a temperature higher than their melting point to analyze the properties of PCMs at liquid state. According to [28] the enthalpy of fusion is calculated by:

$$\Delta H = \int_{t1}^{t2} \emptyset \, dt = \int_{t1}^{t2} \frac{dH}{dt} \, dt \quad (10)$$

where $\phi$ corresponds to the heat flow in mW, $t_1$ y $t_2$ are the time where the analysis was carried out. Besides, the method of determination of heat capacity used in this research is known as direct method, and the principle is derived from the equation that defines the specific heat capacity as:

$$Cp = \frac{dH}{dT}\frac{1}{m} \quad (11)$$

where Cp is the specific heat capacity of the sample, m the mass of the sample, dH/dT is the increase in the enthalpy of the sample when the temperature is increased.The numerator and denominator are differentiated with respect to time. The rate of change of enthalpy (dH/dt) is equal to the DSC heat flow ($\phi$). The rate of change of temperature (dT/dt) of the sample is the heating rate ($\beta$s) that the analyst has settled in the heating rate program. In this way, it is established that:

$$Cp = \frac{dH}{dT}\frac{1}{m} \rightarrow Cp = \frac{dH/dt}{dT/dt}\frac{1}{m} \rightarrow Cp = \frac{\emptyset}{\beta s}\frac{1}{m} \quad (12)$$

# 3. RESULTS AND DISCUSSION

## 3.1. Mixture 54% $KNO_3$ – 46% $NaNO_3$ (System A)

Figure 1 (a) presents the DSC thermogram obtained for the 54% KNO3-46% NaNO3 mixture. The Y-axis represents the heat flow in mW, and the X-axis shows time in minutes. Furthermore, it is important to mention the heat flow direction in the thermogram, for an endothermic process the peak of the curve is concave, so it means the peak is located under the baseline curve. On the other hand, an exothermic process exhibits a peak located above of the baseline curve. The thermogram shows a heat flow increase due to an endothermic process that begins at 107 °C and ends at 119 °C with a peak temperature recorded at 111 ° C. This first endothermic process corresponds to a solid-solid state transition (S-S) of one of the system components, which according to the literature and recorded temperature this thermal phenomenon corresponds to a transition produced by the KNO3. Additionally, a second endothermic process occurs and displays a noticeable peak with a sharp slope, which in comparison to the first peak the heat flow is twenty times bigger. Therefore, this last section represents a solid-liquid state change of the eutectic mixture, starting at a temperature of 221 ° C and ending at 227 ° C with a peak temperature of 223 ° C. The section of the PCM curve where heat flow is constant (slope equal to zero) as the temperature increases belongs to the zone of sensible heat absorption. As time passes by, the slope increases rapidly to reach an infinite value. This zone corresponds to the absorption of heat by a phase change process, and it occurs at constant temperature as it is expected. A shift of slope denotes the melting temperature of the PCM, and the process continues until heat flow attains a maximum value, which is the point where slope returns to zero. This last point indicates that the phase change process has reached its maximum energy requirement in order to melt all the PCM. Since this point, the slope changes again to achieve the level of the sensible heat line, meaning the end of the melting process. Once the curve of heat flow vs. time is obtained, the information is analyzed with the DSC software in order to determine the thermophysical properties of the PCM.

The analysis was carried out by integration of the curve peaks to obtain the enthalpy of fusion, and the starting and ending points of the phase change. It was determined that the 54% KNO3-46% NaNO3 mixture has a melting temperature of 221 ° C, an enthalpy of fusion between 99-111 kJ/kg and a heat capacity for solid state of 1.22 kJ / kg K. The experimental results present good agreement with values reported in the specialized literature, which states that the melting point and enthalpy of fusion for the 54%KNO3- 46%NaNO3 mixture are 222 °C and 108-110 kJ/kg, respectively [8] [10]. It can be seen there are slight differences between both the experimental and theoretical values, which are caused mainly by these factors:

- Purity of the materials used in the experimental tests and those referenced in the literature.
- Integration method of the curve peaks, the type of the base line and the heating rate.

Figure 1 (b) shows the THM curve of the 54% KNO3-46% NaNO3 mixture (D system, see Table 1), where the initial and final temperatures of the liquid- solid phase change process are 210 °C and 231 °C respectively. These values were determined by calculating the first and second inflection points of the THM curve, getting an average value of 220.5 °C. It is important to mention that the specific heat of the mixture in the liquid phase was not calculated by THM because this process requires the reference substance to have a boiling point higher than PCM's melting point, but glycerin boils at 260°C, and it cannot be used for this purpose On the other hand, the computed enthalpy of fusion of the eutectic KNO3-NaNO3 mixture has an average value of 110.2 kJ/kg, while the specific heat of the solid phase is equal to 1.27 kJ / kg K. When comparing these values with the results of DSC, both methods differ by 1°C (0.45%) for the phase change temperature, 2 kJ/kg for melting enthalpy (1.85 %), and 0.055 kJ/kg.K for heat capacity. These results clearly indicate good agreement with the values reported by the literature and it confirms that the mixture of system D has a great potential as PCM for applications in latent heat energy storage due to its high heat of fusion and high density.

### 3.2. Mixture 40% $KNO_3$- 60% $NaNO_3$ (System B)

Figure 2 (a) shows the DSC thermogram of mixture B (solar salt, see Table 1). As in system A, this mixture has an S-S transition peak at a temperature of 127 °C with an enthalpy value of 25 kJ/kg, which is very close to the value in system D. Furthermore, it can also be seen a second peak corresponding to the S-L phase change process that initiates at 220.9 °C and finishes at 253 °C, and it has a peak temperature of 225.98 °C, a heat capacity of 1.33 kJ/kg K for the solid phase, an enthalpy of fusion of 142.33 kJ/kg. Enthalpy was obtained by peak integration using an Integral Tangential baseline unlike system D where a Spline base line was used. Figure 2 (b) depicts the THM curve of system E (solar salt) where the first inflection point indicates that the onset of solidification is at 237 °C. However, it can be seen that the transition between liquid and solid phase is less evident; as result of this, the second inflection point occurs at a temperature under the melting point reported by the literature. To counterbalance this problem, the temperature at which the phase change process ends was determined by assuming that the average value between this unknown temperature and the first inflection point is equal to the melting temperature indicated by the bibliography. With this procedure, the final temperature of the phase change is equal to 203 °C. The enthalpy of fusion of system E, calculated with the THM is equal to 138.9 kJ/kg, whereas the specific heat of the solid phase is 1.38 kJ/kg K. The literature [13] [12] [14] establishes that the theoretic values for system E are 220 °C for melting temperature, 142 kJ/kg for phase change enthalpy, and 1.47 kJ/kg.K for the heat capacity of the solid phase When comparing the results obtained by both DSC and THM methods with those reported in the bibliography mentioned above, it was found that the difference between the DSC results and the literature values differ by 0.33 kJ/kg (0.23%) in the enthalpy of fusion; whereas with the THM there is a variation of 3.01 kJ/kg (2.11%).

### 3.3. Mixture 57% $KNO_3$ -43% $NaNO_2$ (System C)

Figure 3(a) shows the DSC thermogram of the system C. This mixture exhibits an onset point of 137 °C, a peak temperature of 143.94°C, and an endset point of 144.54 °C. The onset represents the beginning of the melting process, and as mentioned before the peak temperature denotes the entire solid has been converted into liquid. The enthalpy was obtained by integration of the peak using an Integral Tangential baseline, and the analysis showed that the enthalpy of fusion of this mixture is equal to 98.6 kJ/kg with a heat capacity of 1.1 kJ/kg K. Figure 3(b) displays the THM curve of system C, and it can be perceived that solidification starts at 146 °C and it ends at 140°C, and the transition from liquid to solid through THM analysis has a enthalpy of fusion of 94.5 kJ/kg. Specialized literature data for enthalpy of fusion and melting point of this mixture are 97 kJ/kg and 145 °C respectively. When comparing both experimental results and specialized bibliography, the melting

point´s deviation percentage of DSC is 5% and THM is 0.68%. As it can be seen, DSC deviation is higher than THM because the purity of $NaNO_2$ used in this analysis was strictly classified as food grade product, for it incorporates impurities that can restrict accurate results due to the DSC sensibility. These impurities can be detected in the thermogram when a non-homogenous heat flow curve is obtained as Figure 3(a) shows, for the melting point is influenced by these impurities as well. On the other hand, the enthalpy of fusion's deviation percentage of DSC is 1.65%, and THM is 2.57% in regard to bibliography.

### 3.4. Mixture 53% $KNO_3$ – 40% $NaNO_2$ – 7% $NaNO_3$ (System D)

Figure 4(a) displays the thermogram of system D, which is known as Hitec in the industrial field. The DSC analysis indicates that melting of Hitec starts at 138°C, and it has a peak temperature of 145.20°C with an endset of 148.31°C. The peak integration was performed by using an Integral Tangential baseline, and it was obtained an enthalpy of fusion of 120 kJ/kg. Furthermore, it can be noticed there is a peak before the melting temperature, as in systems A and B. This effect was mentioned before, and it has an onset point of 123°C, peak temperature of 126°C, and an endset of 129°C. Additionally, this peak was integrated using an Spline baseline obtaining an enthalpy of 117.20kJ/kg. It is important to mention that Hitec is a ternary system, and achieving an homogeneous composition of the sample used in DSC analysis is a challenging task due to the size of the sample, for the result of enthalpy of fusion compared to the bibliography has a notable difference. In contrast, THM analysis offers better results due to the large size sample that this method requires. THM curve can be seen in figure 4(b), and it displays that solidification starts at 144°C approximately, and it ends at a temperature of 140 °C. The enthalpy analysis exhibits an enthalpy of fusion of 82.4 kJ/kg. By comparing DSC method with the specialized literature, it is established that DSC has a deviation percentage of 2.82% for the melting point, and almost 46.5% for the enthalpy of fusion. Likewise, THM method exhibits an error percentage of 1.41% for melting point, and 3% for enthalpy of fusion.

### 3.5. Heat capacity analysis

The DSC technique has greater accuracy for measuring the phase change enthalpy because it is a standardized method, for it means that uncertainty in the thermal properties is minimal. In contrast, the THM, which is a custom method based on considerations and ideal assumptions, could lead to have results with biggest inaccuracy. However, in this study, the deviation percentage of the phase change enthalpy between the THM and the reference value is 2.1%. This percentage, according to the error theory, is within appropriate limits for its use as a mathematical reference. This means that the results presented here by these analyses are appropriate for any engineering application. The equipment software exhibit the heat capacity for solid and liquid states for each temperature as it is showed in Figure 5, but as it is expected heat capacity remains constant before a phase change phenomenon takes place. Furthermore, there is a third zone that shows variable heat capacity represented by the peaks in the curve, but it has to be mentioned that this zone corresponds to the phase transition process, for those values cannot be considered as heat capacity as well because these peaks do not represent sensible heat absorption. Finally, it is important to mention that the experimental values of heat capacity in the solid state differs from those referred in the specialized literature because in this research it was not used sapphire as reference material to determine heat capacity, for the values presented here for Cp in solid and liquid states could exhibit deviations. The data analysis software of the DSC equipment was used to determine the PCMs solid and liquid heat capacity throughout the direct method mentioned in section 2 by using the equipment software analysis. As shown in Figure 5, this software allows obtaining a second thermogram which displays the heat capacity in the Y-axis and the temperature in the X-axis.

### 3.6. Supercooling effect

Throughout the development of the experiments, it was not observed any evidence of supercooling or superheating effects in the DSC technique. According to [29] inorganic PCMs (such as salt hydrates) can sub-cool significantly, whereas organic PCMs (such as paraffin) do not experience serious supercooling. In [30], it is recognized that salt hydrates such as $CaCl_2*6H_2O$, $Na_2SO_4*10H_2O$, $Na_2HPO_4*$ $12H_2O$, $NaCO_3*10H2O$, and $Na_2S_2O_4*5H_2O$ are the most susceptible materials to present supercooling. Based on the literature mentioned above, the mixtures analyzed in this research are not prone to show supercooling effects. Moreover, in this research, the salt mixtures were dehydrated using a stove and a silica gel drier to avoid any interference due to water's presence during analyses.

## 4. CONCLUSION

The main contribution of this research is to validate the T-History method through the application of the differential scanning calorimetry technique (DSC) by a qualitative and quantitative comparison of thermophysical properties of medium temperature phase change materials by applying both techniques simultaneously. Moreover, this research proposes the replacement of water by glycerin as reference material in the THM for the thermal analysis of PCMs with melting point up to 230 ° C. The results obtained by both methods present outstanding agreement with the values established in the specialized literature. Similarly, it can be settled that both methods are complementary, being the DSC the ideal method to determine the melting temperature because of its high sensitivity and technological development. However, by using a sample much bigger than the DSC method, the THM allows to obtain results at bulk level that point out how PCMs will behave in actual operating applications. Another contribution of this paper is the validation of glycerin as reference material for thermal analysis of PCMs through the THM. In this aspect, glycerin is a suitable material than can be used in THM to characterize PCMs of medium range temperature because the thermophysical properties analyzed with THM using glycerin as reference exhibit generally minimal errors when they are compared to the literature and the results of DSC technique. The only exception occurs with the mix 53% KNO3- 40% NaNO2- 7% NaNO3 (Hitec) where it was not possible to apply the DSC method to determine the solid and liquid specific heat, and the value of the phase change enthalpy was concordant neither with literature nor THM. This problem takes place because Hitec is a ternary system, and achieving good homogeneity using minimum sample size that DSC requires becomes very difficult. Besides, an asymmetrical curve of heat flow was obtained, for the sensible heat analysis of the solid phase cannot be done because heat flow is not homogenous.

It is important to comment that in [31] it is exhibited the phase diagram of $KNO_3$ – $NaNO_3$ mixtures which confirms that system B (see table 1) has the highest value of melting point and Hm. Finally, it was determined that the THM method using glycerin instead of water is valid and complementary to DSC for thermal analysis of medium temperature PCMs.

## ACKNOWLEDGMENTS

The authors would like to acknowledge the National Institute of Energy Efficiency and Renewable Energies (INER) for supporting this research and providing the DSC equipment to carry out the experimental thermal analysis.

**Table 1**. Composition of mixtures for thermal analysis

| System | Component | Composition (% wt) |
|---|---|---|
| **A** | $KNO_3$ | 54 |
| | $NaNO_3$ | 46 |
| **B** | $KNO_3$ | 40 |
| | $NaNO_3$ | 60 |
| **C** | $KNO_3$ | 57 |
| | $NaNO_2$ | 43 |
| **D** | $KNO_3$ | 53 |
| | $NaNO_2$ | 40 |
| | $NaNO_3$ | 7 |

**Table 2**. Experimental and theoretical thermophysical properties of medium temperature PCMs

| System | Procedure | $T_m$ $(°C)$ | $H_m$ $(kJ/kg)$ | $Cp_s$ $(J/kg \cdot K)$ | $Cp_l$ $(J/kg \cdot K)$ |
|---|---|---|---|---|---|
| **A** | THM<br>DSC<br>LITERATURE | 220.5<br>221<br>222[8]<br>227[9] | 110.2<br>99-111<br>110[8]<br>108[10] | 1270<br>1215<br>1100-1300[11] | 1180<br>1230<br>- |
| **B** | THM<br>DSC<br>LITERATURE | 237<br>220.9<br>220[12] | 138.9<br>142.3<br>142 [13] | 1380<br>1330<br>1640[13]<br>1470[14] | -<br>1420<br>1490[14] |
| **C** | THM<br>DSC<br>LITERATURE | 146<br>137<br>145[15] | 94.5<br>98.6<br>97[16] | 1275<br>1200<br>1180[16] | 1700<br>1664<br>1740[16] |
| **D** | THM<br>DSC<br>LITERATURE | 144<br>138<br>142[17] | 82.4<br>117.20<br>80[17] | 1110<br>-<br>1340[18] | 1520<br>-<br>1560[19]<br>1560[14] |

**Figure 1. a)** DSC curve for system A**; b)** THM curve for system A

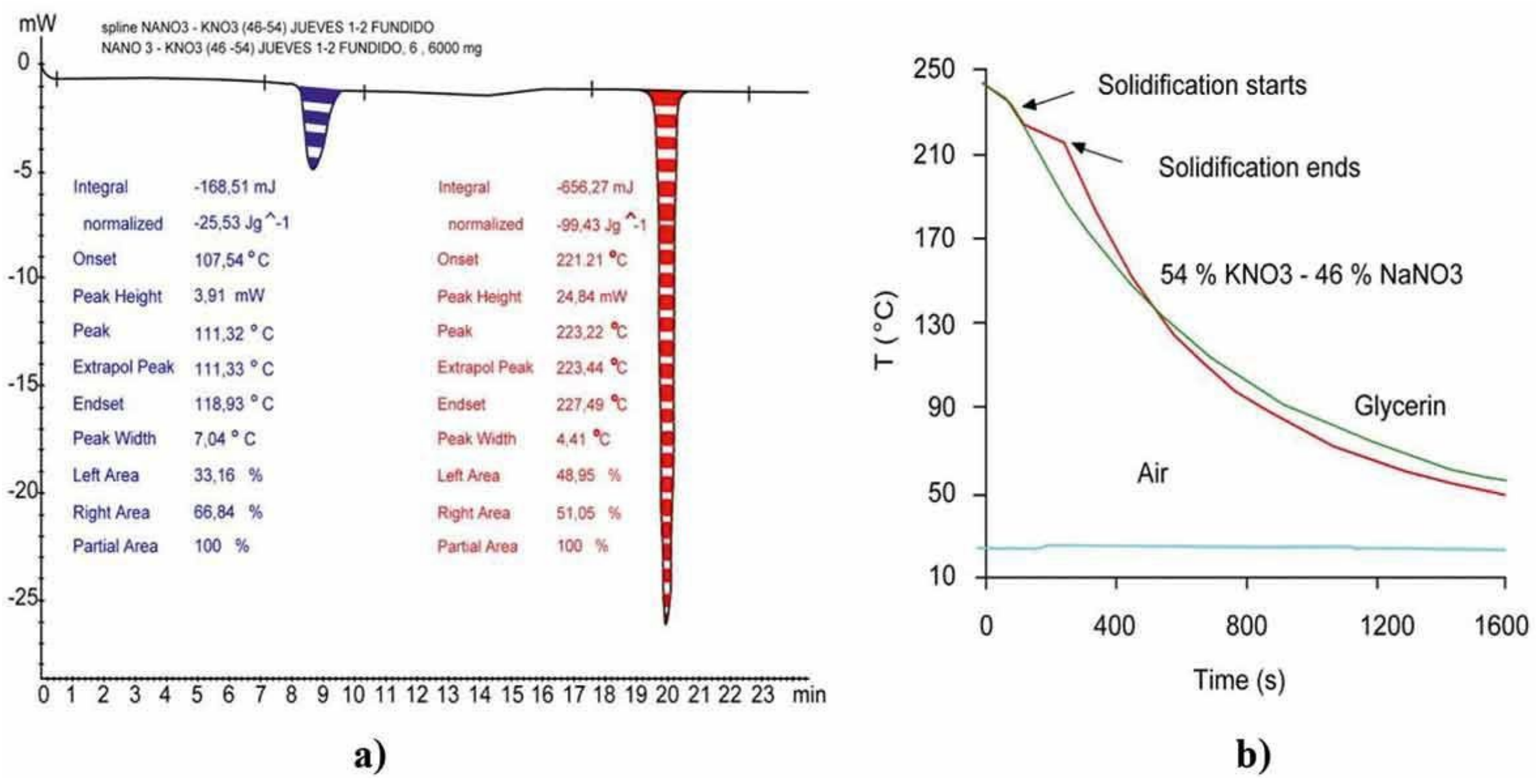

**Figure 2. a)** DSC curve for system B**; b)** THM curve for system B

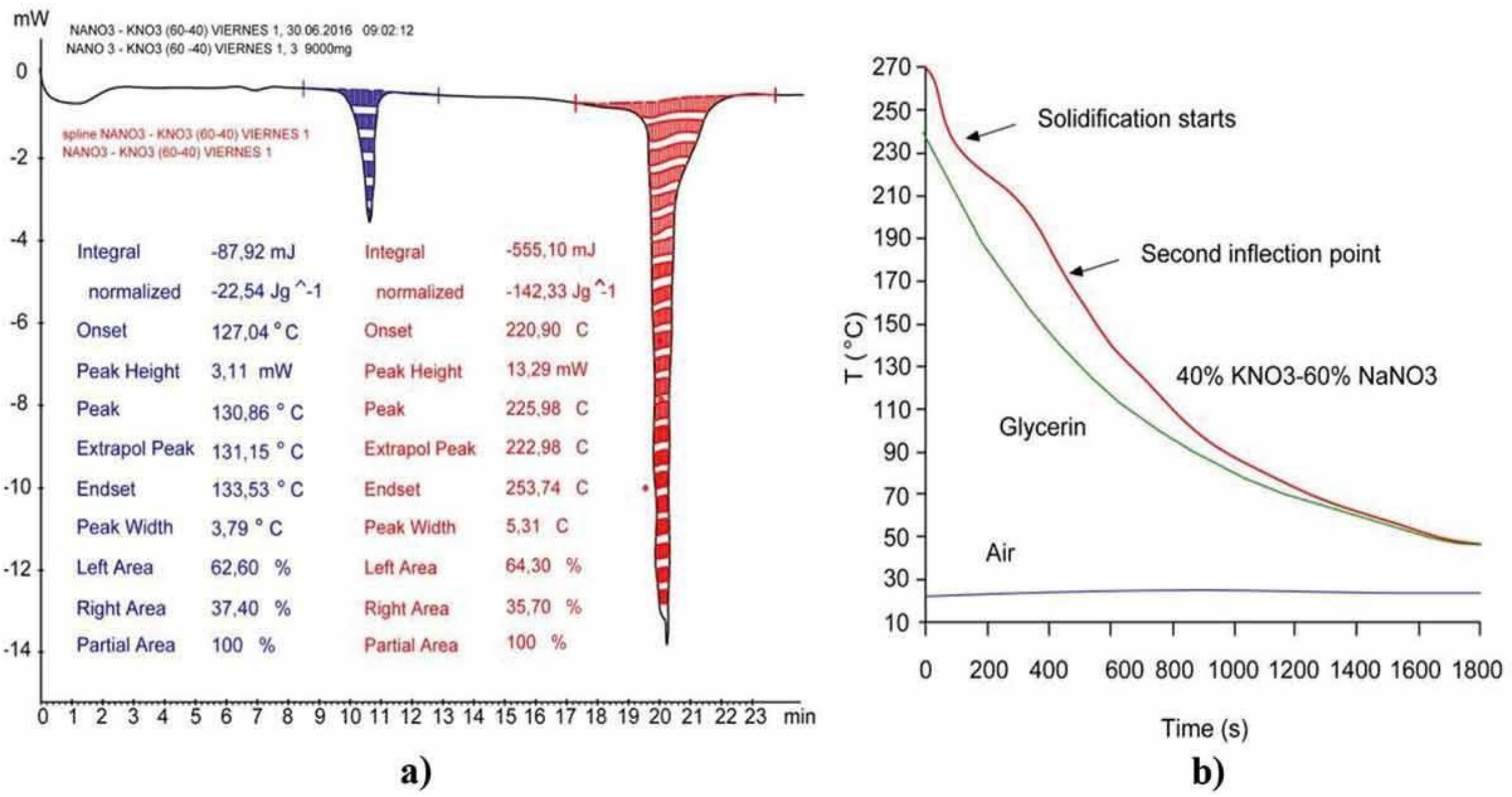

Figure 3. a) DSC curve for system C; b) THM curve for system C

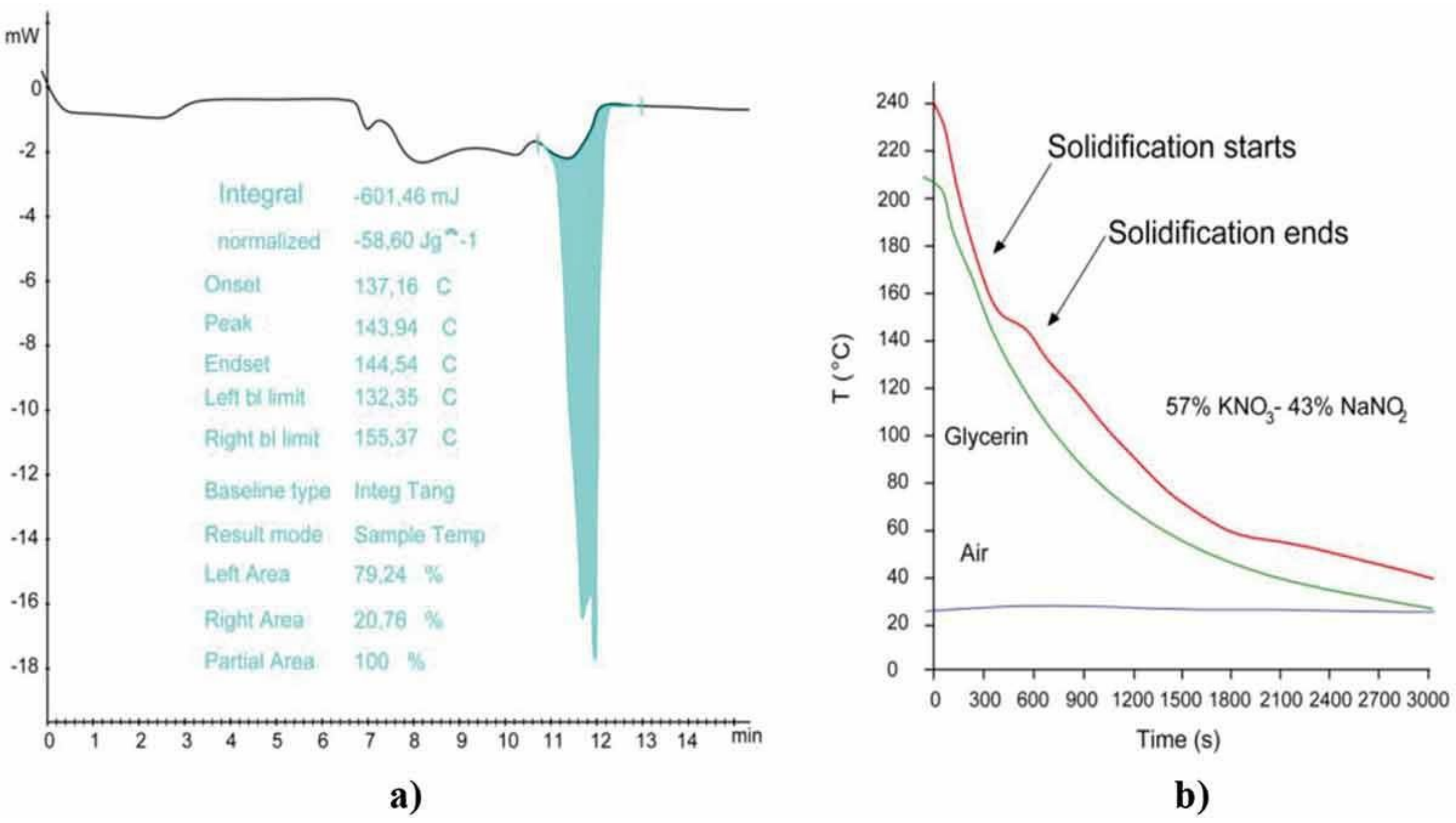

Figure 4. a) DSC curve for system D (Hitec); b) THM curve for system D (Hitec)

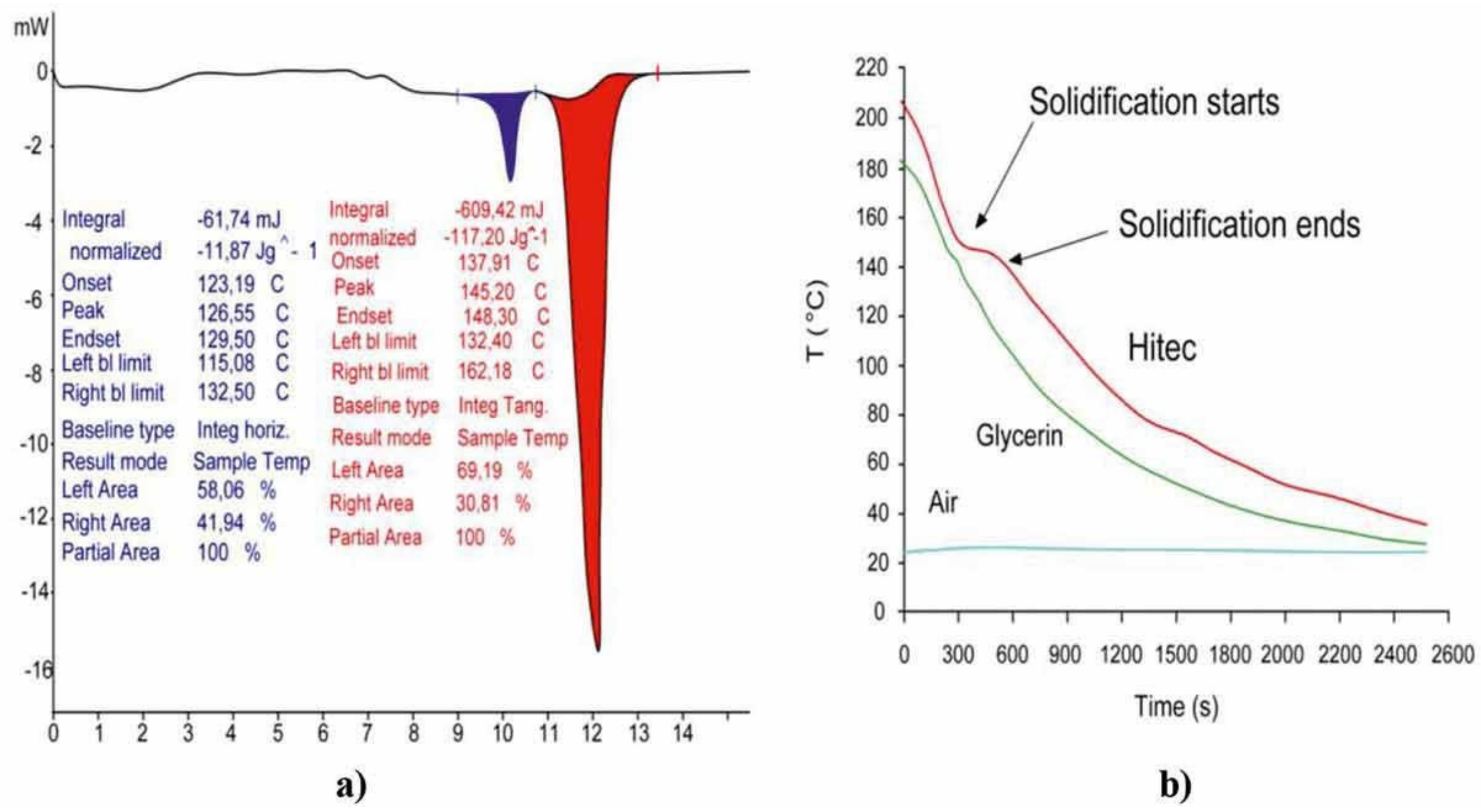

**Figure 5.** DSC thermogram of heat capacity vs temperature for system A

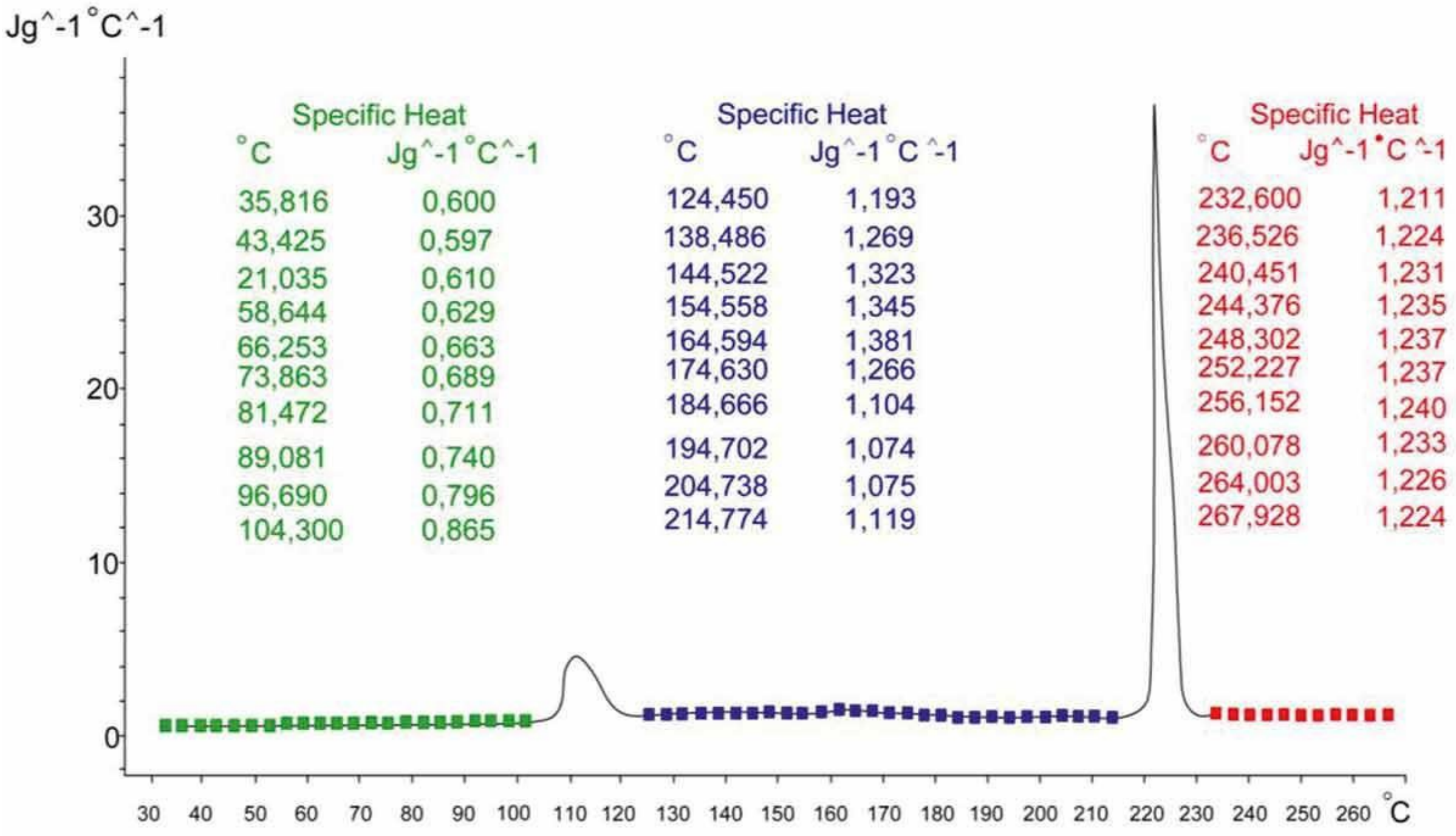